# Pauli blockade of the electron spin flip in bulk GaAs

A. Amo* and L. Viña

*Departamento de Física de Materiales, Universidad Autónoma de Madrid, 28049 Madrid, Spain*

P. Lugli

*Lehrstuhl für Nanoelektronik, TU München, Arcistrasse 21, D-80333 München, Germany*

C. Tejedor

*Departamento de Física Teórica de la Materia Condensada Universidad Autónoma de Madrid, 28049 Madrid, Spain*

A. I. Toropov and K. S. Zhuravlev

*Institute of Semiconductor Physics, Pr. Lavrentieva, 13, 630090 Novosibirsk, Russia*

By means of time-resolved optical orientation under strong optical pumping, the *k*-dependence of the electron spin-flip time ($\tau_{sf}$) in undoped GaAs is experimentally determined. $\tau_{sf}$ monotonically decreases by more than one order of magnitude when the electron kinetic energy varies from 2 to 30 meV. At the high excitation densities and low temperatures of the reported experiments the main spin-flip mechanism of the conduction band electrons is the Bir-Aronov-Pikus. By means of Monte-Carlo simulations we evidence that phase-space filling effects result in the blocking of the spin flip, yielding an increase of $\tau_{sf}$ with excitation density. These effects obtain values of $\tau_{sf}$ up to 30 ns at $k \approx 0$, the longest reported spin-relaxation time in undoped GaAs in the absence of a magnetic field.

## I. INTRODUCTION

The electron spin-relaxation processes in direct gap semiconductors have attracted the attention of the solid-state physics community for the past three decades. This interest has been renewed recently due to the prospective development of spin based devices for the storage, transport and processing of information.[1] Optical orientation has proven to be an extremely powerful tool for the manipulation and study of the electron-spin degree of freedom in direct gap semiconductors:[2] in particular, the spin relaxation mechanisms have been thoroughly studied in III-V materials, such as GaAs, both bulk and low-dimensional.

The theoretical basis underlying the physics of the electron spin-flip processes was established in the late 1970's (a review can be found in Ref. 2), and its understanding is still a very active field of research.[3-5] These theoretical studies have been performed mainly considering doped systems. They have composed a very detailed map of the electron spin-flip relaxation mechanisms under very different conditions of material composition, temperature, and doping density,[4,6,7] and have successfully explained and predicted many of the experimental observations in this field.[8] In the last decade, on the quest for the use of the electron spin as a fundamental constituent in *spintronic* devices, many experimental studies have concentrated on the design and development of GaAs based structures with long electron spin-relaxation times,[6, 9-11] which can reach the microsecond scale.[12]

However, despite all the thorough investigations, some fundamental aspects of the physics of electron spin relaxation in semiconductors have been neglected. One of these issues is the electron-momentum ($k$) dependence of the spin-flip processes. This $k$-dependence is of great importance, not only from a fundamental point of view but also for the design of applications that rely on the transport and injection of electrons with a preserved spin state. In these applications, electrons travel some distance in the system with a non-zero momentum, and a precise knowledge of the $k$-dependence of the spin relaxation time may help to improve the designs.[13-15] This is, in general, carefully accounted for in the theoretical derivations of the spin-flip times ($\tau_{sf}$) under different mechanisms,[3, 4, 7] but has remained largely unexplored in experimental works.

Bulk, $n$-doped GaAs is not suitable for such experimental studies since photoinjected electrons may only keep a spin imbalance at the Fermi edge,[6, 9] and therefore the spin-flip times can only be investigated for electrons with momentum corresponding to this Fermi energy. In the case of $p$-doped and undoped GaAs samples, the measurement of the $k$-dependence of $\tau_{sf}$ by optical means should be feasible, but it has simply not been performed (in the experiments available in the literature, only a single electron energy has been studied in each investigated sample[16-18]). In any case, the use of doped samples should be avoided as doping introduces (*i*) extrinsic scattering centers whose strength may vary depending on the dopant and even depending on the sample (compare $\tau_{sf}$ values, for the same doping concentration, in Refs. 19 and 6); and (*ii*) localization centers that are critical in the determination of the spin-relaxation mechanism.[6, 11, 12]

Another fundamental issue that has not been explored until very recently, is the physics of spin-dependent electron many-body processes[20-22] and phase-space filling effects.[23, 24] Due to the difficulties in the theoretical modeling and in the analysis of the experimental results, the spin relaxation mechanisms in the regime where these effects are important are not well known.

In this communication, we present experimental results that shed some light on the two aforementioned issues, i.e., the $k$-dependence of $\tau_{sf}$, and the spin-dependent many-body and phase-space filling effects on the electron spin-flip processes. We have made use of optical orientation techniques in undoped, bulk GaAs under strong pulsed photoexcitation. We present new experimental results on the $k$-dependence of the spin-flip processes, which, to the best of our knowledge, have not been available up to now. Additionally, our experiments yield the largest observed $\tau_{sf}$ in undoped GaAs.

Although the experiments are self-sustaining and give enough evidence of the influence of *Pauli blocking* on the spin-flip processes, we have made use of Monte-Carlo calculations of the photocreated electron and hole populations in the system to get a further insight on the physical microscopic mechanisms governing the electron-hole scattering under our experimental conditions. A phenomenological scaling law is used to relate the direct electron-hole scattering rates obtained from the simulations and the exchange electron-hole scattering rates obtained from the experiments. Even though the experimental results and the calculations are compared through such a simplified law, the simulations support the interpretation of the main spin-relaxation channel occurring via electron-hole interaction and that phase-space filling effects (*Pauli blockade*) are responsible for the partial inhibition of the spin flip, resulting in the observed long spin-relaxation times.

The remaining of the paper is organized as follows: in section II we describe the samples used in the investigation, the experimental setup and the principles of the optical orientation technique that we have employed to measure the $k$-dependent electron spin-flip times; in section III we present the experimental results; in section IV we describe the Monte-Carlo simulations of the electron and hole ensembles, and discuss the experimental results on the light of the simulations; in section V we summarize the most relevant results presented in this paper.

## II. SAMPLES, EXPERIMENTAL SETUP AND OPTICAL ORIENTATION

We have used high purity, nominally undoped GaAs films grown by Molecular Beam Epitaxy, to perform our experiments. The GaAs layers, with a thickness of 2.5 μm, were encapsulated between two thin AlAs layers (25 nm) in order to minimize the effects of surface recombination. The active region showed a reduced residual *p*-type conductivity with a room temperature hole concentration[25] of ~8×10$^{14}$ cm$^{-3}$.

The samples were placed in a cold-finger cryostat, kept at 5 K, and were optically excited with a tuneable Ti:Al$_2$O$_3$ laser, which produced 1.5 ps pulses at an energy of 1.631 eV, far above the low temperature band gap of GaAs (1.519 eV).[26] This off-resonant excitation enabled the study of the photoluminescence (PL) from electrons with large *k*-states. The laser beam was focused on the sample in a 100 μm-diameter spot. The emitted PL was energy- and time-resolved by means of a spectrometer coupled to a streak camera. The images were recorded in a CCD; the time and energy resolution of the overall setup is better than 9 ps and 0.3 meV, respectively. The pump beam was circularly polarized ($\sigma^+$) by an achromatic λ/4 plate, while additional polarization optics enabled the analysis of the PL into its $\sigma^+$ and $\sigma^-$ components.

In order to measure the electron spin-flip times we made use of optical orientation techniques. In an intrinsic semiconductor, the probability for the excitation of electrons from the valence to the conduction band after a $\sigma^+$-polarized non-resonant pump pulse with photon energy $E$ is given by Eq. (1):

$$\alpha^+(E) \propto \left[\left|\langle e|P_{dip}|hh\rangle\right|^2 \cdot \mu_{e-hh}^{3/2} + \left|\langle e|P_{dip}|lh\rangle\right|^2 \cdot \mu_{e-lh}^{3/2}\right]\left(E-E_g\right)^{1/2}, \qquad (1)$$

where $\langle e|P_{dip}|hh\rangle$ ($\langle e|P_{dip}|lh\rangle$) is the electric dipole matrix element for the absorption of a $\sigma^+$ photon and creation of a spin-down (spin-up) electron and a $J_z = +3/2$ heavy-hole ($J_z = +1/2$ light-hole), $E_g$ is the band gap, $\mu_{e-hh}$ ($\mu_{e-lh}$) is the reduced electron and heavy-hole (light-hole) mass, and $J_z$ is the third component of the total angular momentum. Thus, a $\sigma^+$ incident pulse excites both spin-down and spin-up electrons. Taking into account that $\left|\langle e|P_{dip}|hh\rangle\right|^2$ is 3 times greater than $\left|\langle e|P_{dip}|lh\rangle\right|^2$, if valence band mixing effects are neglected,[27] and that the reduced masses $\mu_{e-hh}$ and $\mu_{e-lh}$ are nearly the same when they are averaged in all directions of space, the maximum injected total electron spin imbalance amounts to ~50%.[28]

The photogenerated holes are also spin polarized, but they loose their spin memory in a time scale of 100 fs,[29] much shorter than any characteristic time considered in this work. Therefore, we will assume that holes are not polarized.

After thermalization and energy relaxation of the carriers in the bands, the electrons recombine with the unpolarized holes, and the $\sigma^+$-polarized light emitted at energy $E$ is given by Eq. (2):

$$I(E,\sigma^+) \propto \left[\left|\langle e|P_{dip}|hh\rangle\right|^2 \cdot f_{e\downarrow}f_{hh\uparrow} \cdot \mu_{e-hh}^{3/2} + \left|\langle e|P_{dip}|lh\rangle\right|^2 \cdot f_{e\uparrow}f_{lh\uparrow} \cdot \mu_{e-lh}^{3/2}\right]\left(E-E_g^*\right)^{1/2}, \qquad (2)$$

where $E_g^*$ is the renormalized band gap, $f_{e\downarrow(\uparrow)}$ are the Fermi-Dirac occupations of spin-down (-up) electrons and $f_{hh\uparrow(lh\uparrow)}$ those of $J_z = +3/2$ heavy- ($J_z = +1/2$ light-) holes. The same selection rules apply to the excitation and emission processes. $\sigma^+$ PL will originate from the recombination of spin-down electrons with heavy-holes, and spin-up electrons with light holes in a ~3 to 1 ratio. The $\sigma^-$ emission $I(E, \sigma^-)$ is given by an expression analogous to Eq. (2) with the arrows in the distribution functions in the opposite direction, meaning a change of sign in the spin or $J_z$. Then, due to the just mentioned 3 to 1 ratio in the optical selection rules, electrons mainly recombine with heavy-holes. The degree of circular polarization of the emitted light, at energy $E$, after excitation with a $\sigma^+$-pulse, is given by Eq. (3):

$$\wp(E) = \frac{I(E,\sigma^+) - I(E,\sigma^-)}{I(E,\sigma^+) + I(E,\sigma^-)}. \qquad (3)$$

$\wp$ provides a direct measurement of the imbalance of the two electronic spin populations.

After the pulsed injection of spin-polarized carriers, they thermalize, slowly cool down and progressively flip their spin towards a spin balanced situation of electrons in the conduction band. The maximum value of $\wp(E)$ is obtained at zero delay after the excitation. The spin-flip rate of the electrons $\tau_{sf}(E)$, at a given energy, can be monitored through the time evolution of $\wp(E)$. The decay time of the polarization $\tau_\wp(E)$ is directly connected to $\tau_{sf}(E)$ through:[30] $\tau_{sf}(E) = 2\cdot\tau_\wp(E)$.

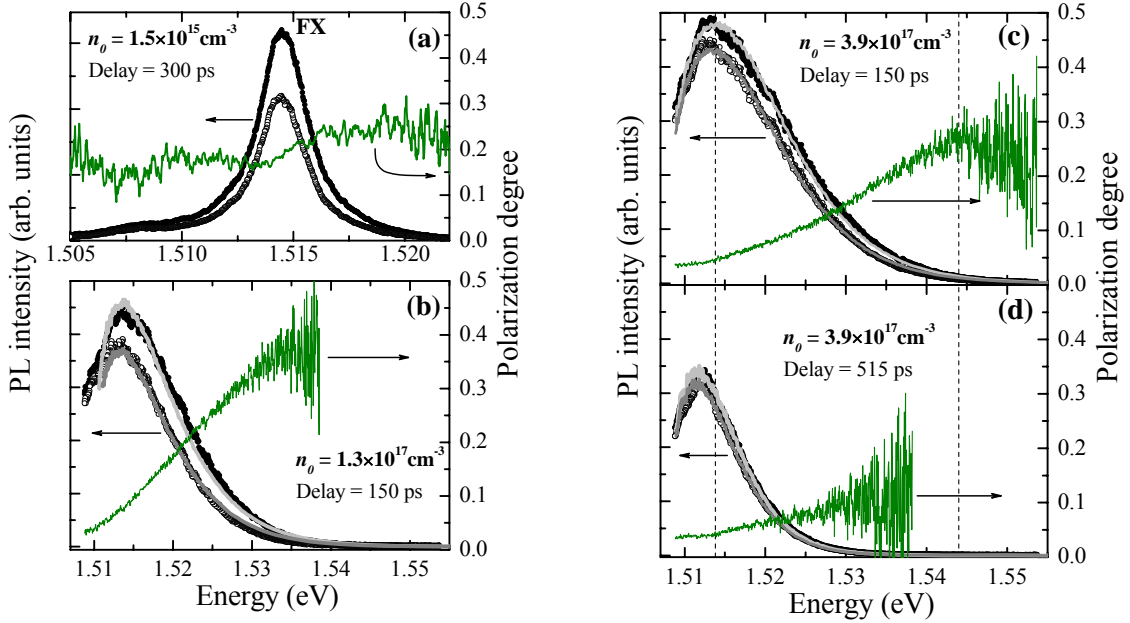

FIG. 1. (Color online) GaAs PL spectra ($\sigma^+$ -closed points, $\sigma^-$ -open points; left scales) and degree of circular polarization (right scales) for different excitation densities $n_0$ and delays after the $\sigma^+$ pulse arrival: (a) very low excitation density, $1.5\times10^{15}$ cm$^{-3}$ at 300 ps; (b) $1.3\times10^{17}$ cm$^{-3}$ at 150 ps; (c) $3.9\times10^{17}$ cm$^{-3}$ at 150 ps; (d) $3.9\times10^{17}$ cm$^{-3}$ at 515 ps. The thick solid lines are fits to the PL as explained in the text. FX denotes the free exciton line.

Let us point out that the aforementioned polarization- and time-resolved PL measurements provide a direct quantification of $\tau_{sf}$, without the possible spurious effects associated to the use of external magnetic fields and/or post-experimental theoretical fittings which are inherent to other techniques, such as those based on the Hanle or Kerr effects. It also avoids other limitations present in techniques like time-resolved photoemission, which just probes the surface of the sample where the localization of carriers destroys any possibility of accessing the information on the electron momentum.[31]

### III. EXPERIMENTAL RESULTS

Figure 1(a) shows the GaAs PL spectrum 300 ps after the arrival of the laser pulse at low excitation density ($1.5\times10^{15}$ carriers$\times$cm$^{-3}$). At such low power, the photoexcited electrons and holes mainly form free excitons. Only those excitons with center of mass momentum $K \approx 0$ can couple to light. Hence, the spectral characteristics of the free exciton line primarily arise from the homogeneous nature of the resonance and from exciton dephasing processes.[32, 33] A low energy shoulder (1.508 eV) coming from electron-acceptor recombination[34] can also be distinguished in the figure. Due to the homogeneous origin of the free-exciton line, no spectral dependence of $\wp$ is expected across the resonance, as observed in Fig. 1(a): $\wp$ remains almost constant along the free exciton emission, and in particular at its high-energy side with a value of 0.21±0.04. The slight abrupt jump ($\Delta\wp \leq 0.05$) at the center of the line originates from a splitting between the $\sigma^+$ and $\sigma^-$ components of the PL (~0.1 meV), due to interexcitonic interactions.[35]

At high excitation densities, above $1.2\times10^{17}$ cm$^{-3}$, screening of the carriers leads to the formation of an electron-hole plasma.[36] In this case, the $\sigma^+$ photoexcitation creates two ensembles of electrons with a high excess energy in the conduction band (with down- and up-spin in a ~3 to 1 ratio as discussed earlier). In a time scale of the order of 1 ps, due to strong carrier-carrier and carrier-phonon interactions, each population thermalizes conforming broad Fermi-Dirac distributions in the band[37, 38] with a temperature well above the lattice temperature. Simultaneously, an analogous process for the depolarized holes leads to the achievement of a thermal distribution also in the valence band. The carrier distributions then slowly cool down towards the lattice temperature through carrier-phonon interaction. For such high carrier densities in the system, where exciton formation is hindered,[36] electrons and holes with any $k$ can radiatively recombine, as long as the total electron-hole pair momentum is close to *zero* (only *vertical* transitions between the bands are allowed). Electrons with finite kinetic energy can then recombine at energies above the gap. In this situation the PL lineshape does not originate from the homogeneous character of the resonance (as in the case of excitons). The kinetic energy $E_{k-e}$ and the momentum $k$ of the electrons that recombine at an emission energy $E$ are related by:

$$E_{k-e} = \frac{\hbar^2 k^2}{2 m_e} = \frac{m_h}{m_h + m_e}\left(E - E_g^*\right), \quad (4)$$

where $\hbar$ is Plank's constant over $2\pi$, $m_e$ is the electron effective mass, and $m_h$ is the heavy- or light-hole mass depending on the kind of hole with which the electron recombines.

Figures 1(b)-(d) show the PL spectra for two excitation densities in the electron-hole plasma regime, at different delays after the excitation pulse arrival. Very broad emission from the plasma is observed (notice the *x*-scales). The graphs show the widest spectral window allowed by our setup; the central detection energy was chosen in order to cover the high energy tail, which contains all the information about the electron populations. The large amount of injected carriers produces a renormalization of the band gap, due to exchange and correlation effects.[39] Band-gap renormalizations as large as 25 meV have been reported in similar systems under analogous conditions.[40]

As clearly seen in Fig. 1(b), the degree of circular polarization shows a strong spectral dependence, which results from the spin imbalance of the two spin electron populations. At short delays after excitation, the occupation of electron states with low $E_{k-e}$ (emission energy close to the renormalized band gap, 1.508 eV) is very similar for both electron spin populations, resulting in very low values of $\wp$. However, for higher $E_{k-e}$ there are progressively more spin-down than spin-up electrons, yielding higher polarization degrees, which approach values of 0.4 for the lowest initial carrier density at high energy. At larger densities [Fig. 1(c)] an analogous spectral dependence of $\wp$ can be observed but with smaller values. This decrease of $\wp$ with power can originate from a broader initial distribution of carriers together with a reduction of $\tau_{sf}$ with $E_{k-e}$ (see below), with increasing excitation density, and/or from spectral-hole burning effects which are more important for electrons excited from the heavy-hole than from the light-hole band.[41] At latter times [Fig. 1(d)], the spin flip processes, which tend to balance both populations, produce an overall polarization degree decrease. Nonetheless, the monotonically increasing behavior of $\wp(E)$ with emission energy is preserved.

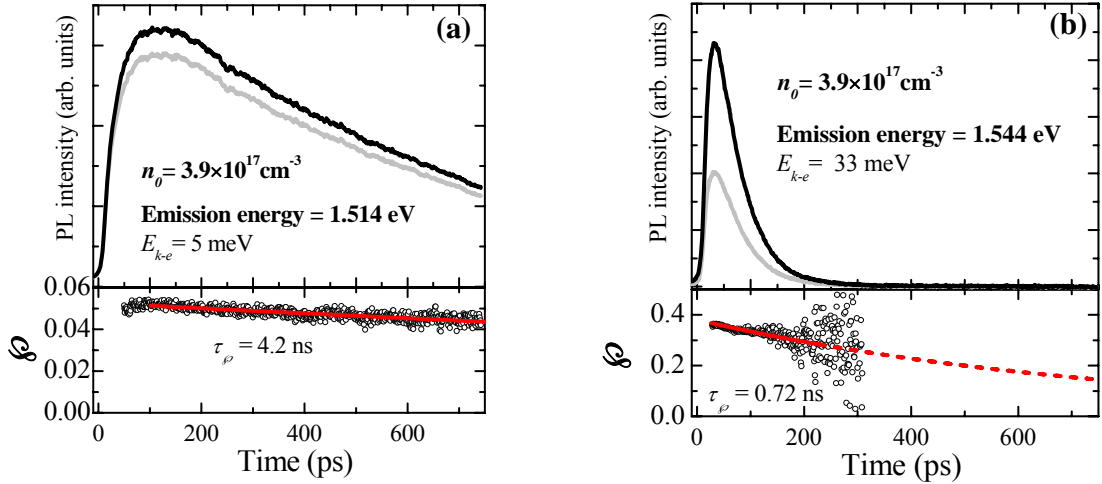

FIG. 2. (Color online) PL intensity (upper panels) of the $\sigma^+$ (black line) and $\sigma^-$ (grey line) components, after $\sigma^+$ excitation, for the highest excitation density ($3.9\times10^{17}$ cm$^{-3}$), at two emission energies [denoted by vertical lines in Figs. 1 (c) & (d)]: (a) 1.514 eV ($E_{k-e}$ = 5 meV); (b) 1.544 eV ($E_{k-e}$ = 33 meV). The lower panels show the corresponding degree of circular polarization. The lines are fits to a exponential decay function, with polarization decay times of 4.2 ns in (a) and 0.72 ns in (b). Note the different vertical scales in (a) and (b).

We can now focus on the PL dynamics at different emission energies. Figure 2 depicts the time evolution traces of the $\sigma^+$ and $\sigma^-$ luminescence for the highest investigated excitation density at two different emission energies [corresponding to the vertical lines in Figs. 1(c) & (d)]. Both rise and decay dynamics are very different in the two cases as a consequence of the relaxation and cooling dynamics of the electron ensembles. At low (high) energies the PL evolution reflects the radiative recombination and the filling (emptying) of electronic states. The cooling process of the electron populations from high- to low energy states results therefore in a slow (fast) dynamics at low (high) energies.[31, 36]

In the lower panels of Fig. 2 the time evolution of $\wp$, extracted from the upper panels traces, is presented. $\wp$ decays with time and $\tau_\wp$ can be obtained by fittings to monoexponential decay functions.

Figure 3(a) depicts the electron spin-flip time $\tau_{sf}$ obtained from $\tau_\wp$, for different excitation densities as a function of electron kinetic energy in the conduction band. To obtain the kinetic energy we have used Eq. (4) assuming that the emission energy comes from electron and heavy-hole recombination. The renormalized band-gap energy has been obtained from the fits to the PL that will be discussed in section IV. $\tau_{sf}$ increases with excitation density and decreases with increasing electron kinetic energy. Values of $\tau_{sf}$ up to 26 ns are obtained for low $E_{k-e}$ at the highest density. These are the longest spin-relaxation times reported in a nominally undoped GaAs sample, and of the same order than those reported for lightly doped $n$-type GaAs.[19] As we will discuss in the following section, the observation of such long $\tau_{sf}$ is related to the *Pauli blockade* of the spin-flip processes for electron states with occupations close to 1.

## IV. MONTE-CARLO SIMULATIONS AND DISCUSSION

The three main spin relaxation mechanisms in bulk zinc-blende semiconductors are the Elliot-Yaffet (EY),[42] D'yakonov-Perel (DP)[43, 44] and Bir-Aronov-Pikus (BAP).[45] EY arises from the presence of spin-orbit coupling during electron collisions with phonons and impurities, and it is important only for narrow-gap semiconductors or very low temperatures.[4] DP originates from the spin splitting caused by the absence of inversion symmetry. The splitting can be modeled as originating from an effective internal magnetic field in the direction and magnitude of $k$, giving rise to a $\tau_{sf}$ inversely proportional to the electron scattering time.[43, 44] On the other hand, BAP relies on the electron-hole exchange interaction, as holes randomize their spin very fast due to valence band mixing, and its corresponding $\tau_{sf}$ is proportional to the electron-hole scattering time.[45]

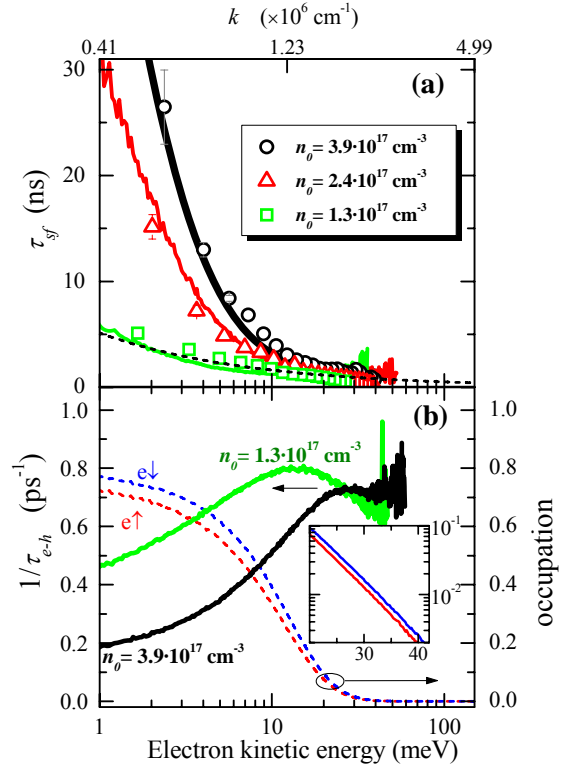

FIG. 3. (Color online) (a) Measured spin-flip time as a function of the electron kinetic energy for initial excitation densities of $3.9\times10^{17}$ cm$^{-3}$ (circles), $2.4\times10^{17}$ cm$^{-3}$ (triangles) and $1.3\times10^{17}$ cm$^{-3}$ (squares). Solid lines compile the fitted spin-flip time as discussed in the text. The dashed line corresponds to the non-degenerate case with a hole density of $1.43\times10^{17}$ cm$^{-3}$, which corresponds to the mean hole density during the PL lifetime for the highest excitation case ($n_0 = 3.9\times10^{17}$ cm$^{-3}$). (b) Occupation of the electron states with spin-down and -up (dashed lines), as well as the total electron-hole scattering rate (solid dark line) for the excitation and time delay shown in Fig. 1(c). The solid light line depicts the scattering rate for the conditions of Fig. 1(b). The inset presents a zoom of the electron occupations in order to clearly show the electron spin imbalance in the 20-42 meV electron kinetic energy range. The non-linear upper scale in (a) corresponds to the momentum of the electrons.

For temperatures above ~4 K and very weak photoexcitation, DP is the dominant mechanism in *n*-GaAs, and in *p*-GaAs for doping concentrations below $10^{16}$ cm$^{-3}$. At higher acceptor concentrations, the main spin-relaxation mechanism is BAP.[4] Theoretical calculations predict that in undoped samples under weak photoinjection, the main spin relaxation mechanism is DP.[4] We are interested in a situation of strong optical pumping in an undoped semiconductor, which has not been explored in detail.[17] Under this circumstance, a high electron concentration would favor a spin relaxation mechanism based on DP, but, on the same ground, a high hole concentration would result in the enhancement of the BAP mechanism.[4, 6, 9] However, if we extrapolate the results for electron spin relaxation obtained in doped samples, where the BAP relaxation rate in *p*-type materials is much stronger than the DP in *n*-type for the same doping concentrations,[4, 11] we expect the BAP mechanism to dominate the electron spin relaxation in undoped samples under strong pumping. The BAP mechanism is so efficient in flipping the electron spin as compared to DP that even in *n*-doped samples the presence of photoexcited holes has been proved to shorten the spin lifetime due to the efficient electron-hole scattering (see Ref. 46 and the inset of Fig. 2 in Ref. 6). Additionally, calculations of Fishman *et al.*[18] and Maialle[3] show that for hole concentrations similar to the ones photoinjected in our system and electron kinetic energies below 100 meV, the spin relaxation rate associated to the BAP mechanism is up to 3 orders of magnitude greater than that associated to the DP mechanism.

In strongly excited systems, on the top of the intrinsic spin-relaxation mechanisms, phase-space filling effects can be of great importance. Spin-flip rates can be highly influenced by the occupation of the final electron state. In the framework of a spin relaxation fully dominated by the BAP mechanism, and in order to explore the influence of the phase-space filling effects, we have performed Monte-Carlo simulations of the carrier populations that evaluate the electron-hole scattering rates ($1/\tau_{e-h}$) under our experimental conditions.

The injected electrons and holes in the Monte-Carlo simulations scatter among themselves and with phonons, conforming thermalized Fermi-Dirac distributions. The carrier-carrier and carrier-phonon scattering processes, which carefully include final-state exclusion effects, are accounted for by using a static multiscreening approach as discussed in Ref. 37. We have used the same material parameters as in Ref. 37, with the simplification of considering only degenerate $\Gamma$ conduction bands, and a heavy-hole valence band, all of them parabolic. The simplification in the use of a heavy-hole valence band only, is justified in our case as the electron–light-hole scattering has a very weak influence on the electron spin-flip time.[3]

To obtain $\tau_{e-h}$ one needs a precise knowledge of the electron and hole distribution functions, $f_{e\uparrow(\downarrow)}$ and $f_{hh}$, for given excitation powers and time-delays after excitation (the spin in the hole distribution can be neglected due to the depolarization of the holes). A fitting of the PL spectra at different delays with $I(E, \sigma^+)$ [Eq. (2)] and the corresponding $I(E, \sigma^-)$, using the density of each type of carriers, its temperatures and the renormalized gap as fitting parameters, yields the time-dependent $f_{e\uparrow(\downarrow)}$ and $f_{hh}$. In the fits, electrons and holes are forced to stay in thermodynamic equilibrium. The thick solid lines in Figs. 1(b)-(d) depict the results of these fits for different excitation powers and delays.

The main role of the Monte-Carlo method is the calculation of the electron-hole scattering rates for each fitted delay. In the simulations, a carrier is randomly chosen from the spin-up/-down electron or hole distributions and made to scatter with electrons, holes or phonons with a probability given by the above mentioned static multiscreening approach. Accounting for the exchange interaction in inter-particle scattering goes beyond the scope and capabilities of the standard Monte-Carlo simulation of the relaxation dynamics of a photoexcited electron-hole distribution. To the best of our knowledge, only very preliminary attempts in such direction have been presented in the literature (see, e.g. Ref. 47). We have therefore considered explicitly only the direct Coulomb-like carrier-carrier interactions, including the effect of the exchange interaction phenomenologically via two fitting parameters, as will be explained in detail later. The electron-hole scattering rates are calculated by integrating during a fixed time interval (5 ps), the number of scattering events between holes and electrons of either spin that are not frustrated by the final state occupation of the scattering partners. The electron and hole distributions are forced to be $f_{e\uparrow(\downarrow)}$ and $f_{hh}$ during the integration interval.

Figure 3(b) depicts $1/\tau_{e-h}$ (thick solid lines) for the scattering of electrons with heavy holes, in the conditions of Fig. 1(b) -light line-, and Fig. 1(c) -dark line-. For the latter case, also the Fermi-Dirac distributions of both electron populations are shown (dashed lines). The scattering rates of low kinetic-energy electrons are considerably smaller than those of high-energy ones, due to the higher occupation of the former (a factor of 4 from $E_k = 25$ meV to $E_k = 1$ meV for $n_0 = 3.9 \times 10^{17}$ cm$^{-3}$). The effect of the electron occupation on $1/\tau_{e-h}$ is also evidenced when comparing the electron-hole scattering rates for different excitation densities: at low $E_k$, $1/\tau_{e-h}$ is more than twice as large for low carrier density [light line in Fig. 3(b)] than for high carrier density [dark line in Fig. 3(b)], where the higher occupation of electronic states inhibits electron-hole scattering.

So far we have discussed the simulations for a time delay of 150 ps. As time evolves, the carrier distributions slowly change due to the radiative recombination and to the cooling of the ensembles. However, our results show that $\tau_{e-h}$ changes only slightly during the PL lifetime. To account for these small changes, we have averaged $\tau_{e-h}$ during the decay time of the PL to reach $1/e$ of its maximum value.

In order to compare our simulation results for the direct electron-hole scattering rate with our experimental results on the spin-flip rate (which is equivalent to the electron-hole exchange scattering rate in the BAP mechanism) we have made use of a scaling relation of the type:

$$\frac{1}{\tau_{sf}(k)} = C \left( \frac{1}{\tau_{e-h}(k)} \right)^{\beta}, \quad (5)$$

where $C$ and $\beta$ are the scaling coefficient and scaling exponent respectively showing no dependence on $k$. Analogous scaling relations can be inferred from calculations on the electron-electron scattering rates including and excluding exchange[48] (see discussion below), and can also be deduced for the case of the correlation and exchange mean energies per electron in an electron-hole plasma in the context of a screened potential approximation.[49]

We have fitted the experimental points shown in Fig. 3(a) with Eq. (5) using the simulated averaged $\tau_{e-h}$ for each excitation density. The fit was performed simultaneously for the three considered excitation densities to obtain the fitting parameters $C$ and $\beta$. The results of the fit is shown in Fig. 3(a) as solid lines, yielding values of $C = 2.05 \cdot 10^{-3}$ and $\beta = 2.81$. The agreement is excellent for the three investigated densities, in particular for the highest two. In the case of lowest excitation power, the injected carrier density ($1.3 \times 10^{17}$ cm$^{-3}$) is very close to the critical density ($1.2 \times 10^{17}$ cm$^{-3}$) for the Mott transition,[36] below which excitons greatly contribute to the PL at energies close to the gap. The excitonic contribution to the polarization dynamics of the PL is not accounted for by our model and may be responsible for the deviations in the fit for the set of experimental points at lowest excitation density, particularly at low electron kinetic energies.

Our scaling model of the direct and exchange spin-flip electron-hole scattering rates given by Eq. (5) is based on the idea that the spin flip is dominated by exchange interaction. Both direct and exchange interactions scale with different powers of the density, so that one can scale exchange as a power of the direct term. This general argument has proved to be valid in calculations of carrier-carrier scattering by Collet in undoped GaAs in a static screening approximation.[48] From the results shown in Fig. 5 of Ref. 48 for the electron-electron scattering rates as a function of density, including and excluding the exchange interaction, a scaling relation between direct and exchange scattering rates analogous to Eq. (5) can be extracted. In that case of electron-electron scattering a scaling exponent close to 3.5 would be obtained, not far from our result of $\beta = 2.81$ for electron-hole scattering.

In order to understand the phase-space filling effects on the spin-flip time, we have calculated $\tau_{sf}$ for the case of non-degenerate holes and empty conduction band. The dashed line in Fig. 3(a) shows $\tau_{sf}$ as derived by Bir et al,[45] and reformulated in Ref. 3 as:

$$1/\tau_{sf}^{non-deg} = N_h \sigma_s \left( 2/\mu_{e-hh} \right)^{1/2} E_{k-e}^{1/2}, \quad (6)$$

where $N_h$ is the hole density and $\sigma_s$ is the spin-flip cross section. For our evaluation of $\tau_{sf}^{non-deg}$ we have used a hole density corresponding to the highest investigated density in our experiments, which is far beyond the assumption of non-degeneracy, but provides a qualitative reference for the effect of phase-space filling on the spin-flip rates.[3] To visualize the effects one has to compare the dashed line with the bold solid line: for low kinetic energies $\tau_{sf}$ is greatly increased with respect to $\tau_{sf}^{non-deg}$ (~8 times, from 3.6 ns to 29 ns for $E_{k-e} = 2$ meV) due to the frustration of the electron spin-flip in highly occupied states (*Pauli blockade*). At higher electron kinetic energies (>20 meV), the occupation is much lower [see Fig. 3(b)] and $\tau_{sf}$ approaches the non-degenerate values. Thus, the *Pauli blockade* of the spin relaxation does not only modify the overall value of $\tau_{sf}$, but it also affects the energy dependence of the spin-flip processes.[3]

The *k*-dependence of the electron spin-flip time reported in Fig. 3(a) shows spin-flip time values and follows trends very close to those calculated by Maialle in bulk GaAs, assuming a BAP spin relaxation mechanism, with *p*-doping concentrations very similar to the photoinjected electron-hole pair densities of our experiments.[3] Our measurements yield slightly higher values of $\tau_{sf}$ caused by a higher *Pauli blockade* of the electron-hole scattering due to the presence of degenerate electron populations (the calculations of Ref. 3 are performed in the absence of electrons in the conduction band, but account for degenerate valence bands). Despite the differences in the system conditions between the calculations of Ref. 3 and the experimental results shown in Fig. 3(a), the good qualitative agreement between the two supports our model of electron spin relaxation through the BAP mechanism under strong *Pauli blockade* for our experimental results.

Let us finally note that the DP mechanism would result in a very different dependence of $\tau_{sf}$ with excitation density. In the DP mechanism, the spin-flip rate is inversely proportional to the electron scattering rate, while in our model both rates are proportional to each other [see Eq. (5)]. Our experiments clearly show an increase of $\tau_{sf}$ with increasing optically pumped carrier density. The simulations obtain an increase of the electron scattering time with density, due to *Pauli blockade* [compare light and dark solid lines in Fig. 3(b)], confirming the validity of our model and discarding the DP mechanism.

## V. CONCLUSIONS

By means of optical orientation techniques the *k*-dependence of the electron spin-flip times in a direct gap semiconductor has been measured. In the case of undoped GaAs under strong photoexcitation, where the densities of free electrons and holes are identical, the main spin-flip mechanism is BAP. By means of a Monte-Carlo simulation, we have evidenced that in this situation of a highly degenerate system, the large occupation of low energy states frustrates the electron spin relaxation, yielding an increase of $\tau_{sf}$ of up to 8 times as compared with a non-degenerate system. This *Pauli blockade* also affects the energy dependence of $\tau_{sf}$, as the occupation of electrons in the conduction band follows a Fermi-Dirac distribution, with the highest occupation for the lowest energy electrons. Additionally, we have found a simple scaling relation between the direct Coulomb and exchange electron-hole scattering rates. All these facts should be taken into account in the design of semiconductor devices based on the transport of electrons with a preserved spin orientation.

## ACKNOWLEDGEMENTS

We thank L. C. Andreani for fruitful discussion. This work was partially supported by the Spanish *MEC* (MAT2005-01388; NAN2004-09109-C04-04), the Comunidad Autónoma de Madrid (S-0505/ESP-0200) and the Russian Basic Research Fund (grant no. 04-02-16774a). A. A. acknowledges a scholarship of the *FPU* program of the Spanish *MEC*.

## REFERENCES AND FOOTNOTES